\documentstyle[12pt]{article}

\begin{document}

\newcommand{\bib}{\bibitem}
\newcommand{\er}{\end{eqnarray}}
\newcommand{\br}{\begin{eqnarray}}
\newcommand{\be}{\begin{equation}}
\newcommand{\ee}{\end{equation}}
\newcommand{\epe}{\end{equation}}
\newcommand{\bea}{\begin{eqnarray}}
\newcommand{\eea}{\end{eqnarray}}
\newcommand{\ba}{\begin{eqnarray}}
\newcommand{\ea}{\end{eqnarray}}
\newcommand{\epa}{\end{eqnarray}}
\newcommand{\ar}{\rightarrow}

\def\r{\rho}
\def\D{\Delta}
\def\R{I\!\!R}
\def\l{\lambda}
\def\D{\Delta}
\def\d{\delta}
\def\T{\tilde{T}}
\def\k{\kappa}
\def\t{\tau}
\def\f{\phi}
\def\p{\psi}
\def\z{\zeta}
\def\ep{\epsilon}
\def\hx{\widehat{\xi}}
\def\na{\nabla}
\begin{center}

{\bf Generalized geometries and kinematics for Quantum Gravity.}

\vspace{1.3cm}
M. Botta Cantcheff\footnote{e-mail: botta@cbpf.br}  

\vspace{3mm}
Centro Brasileiro de Pesquisas Fisicas (CBPF)

Departamento de Teoria de Campos e Particulas (DCP)

Rua Dr. Xavier Sigaud, 150 - Urca

22290-180 - Rio de Janeiro - RJ - Brazil.
\vspace{5mm}

\end{center}

\begin{abstract}
 Our proposal here is to set up the conceptual framework for an eventual
 {Theory of Everything}.

We formulate the arena -language- to build up {\it any} QG. In
 particular, we show how the objects of fundamental theories, such as
 p-branes (strings, loops and others) could be posed in this language. 

\end{abstract}
\section{Introduction}

What are the quantum states of Quantum Gravity (QG)? The main purpose of this paper is to
 find them in the most natural way, in order to obtain a framework  general enough to embody a
  complete theory of Nature, such that the states of the conjectured fundamental
   theories \cite{rev} be particular cases. Rephrasing: we wish to establish a "kinematics for QG". 

Usually, the basis manifold in Quantum Field Theory (QFT) is assumed to be fixed: d-dimensional, differentiable,
 endowed with a metric tensor and a compatible covariant derivative. 

On the other hand, 
 there are various approaches to background-independent
 theories, more or less genuinely
 background independents (loops, strings, branes, posets,
 and so on \cite{rev}). 
 Here, starting from very fundamental
  considerations, we present {\it unifying perspectives
 on these}, which actually generalizes to a
  general scheme for background independence.

In these
  usual approaches to QG, the manifold is thought as
 the picture in large scale of more
   fundamental geometrical structures, which are {\it some type}
 of
 lower-dimensional manifold, embedded in a {\it given} "ambient space";
 they are {\it in
  somewhere}
 (thus, they they appears to be not completely background-independent
 formulations). 

Thus, the natural idea is to adopt the most general point of view; to assume that
 "the states of the spacetime are themselves manifolds, or collections of
 manifolds (multi-manifolds, M-M) \footnote{Or, more weakly, some set of points.}",
 whose topology and dimension are in principle unrestricted. In a M-M state, for instance,
 each component could have different dimension and topological structure.

Furthermore, we shall assume that the physical fields {\it live} (are defined on) this
 generalized geometries . This is in accordance with a  conception of the spacetime where it
  is defined in terms of the phenomenology \cite{pst}.

So to speak, in this work, we propose a generalization of the concept of background
  of a full field theory; in our treatment, they are the space-time states. This
   generalization is motivated by the need of a well-defined quantum theory
   of gravity. This would constitute clearly a formulation based upon a 
   {\it background-independent} fashion, by construction. Moreover, it would be in
    agreement with other conceptual requirements \cite{rov}.

	 This paper is organized according to the following outline:
	 
	 In Section 2, the main assumptions are established and the most general Hilbert
 space for QG is built; next, in Section 3, "embedding-type" structures, which are backgrounds
 embedded in other one, are described; and p-branes (and strings)
 and loops are proposed as examples. Questions related to dynamics for QG are also
 commented.
	 
	 Finally, our concluding remarks are collected in Section 4.

	 \section{QFT and bases of backgrounds: main assumptions.}

In a Hamiltonian Field Theory (FT), the set of degrees of freedom is given by a complete set
 of commuting observables (CSCO) (the fields). For example, in a Klein-Gordon Field Theory,
  the CSOC is the scalar field $\rho(x)$, where $x$ is an element of a spatial Cauchy surface
   $\Sigma(\sim\R^3)$; thus, the CSOC may be expressed by $(\Sigma;\rho(x))$.

In a classical theory, the most general CSCO is given by
 the background, which consists
(in classical physics) of a differentiable basis manifold $M$ with a metric $g_{ab}$
\footnote{And a compatible covariant derivative operator $\nabla_a$, i.e $\nabla_a g_{bc}=0$.},
 plus a collection of smooth fields (the metric is usually thought as other one) $\phi$. The
  field dynamics is currently governed by the approach of gauge theories, while General Relativity (GR)
 describes the background.

If we restrict ourselves to globally hyperbolic space-times, the background structure is fully
 characterized by the geometry of a Cauchy's surface
  $\Sigma$\footnote{$M\sim\Sigma \times \R$.}. Then, let us express by $B$ the set of variables
   of the spatial geometry which characterizes its degrees of freedom (CSCO);
    for instance, topology, dimension, metric, connection, and so on, which {\it a priori} will be
     arbitrary \footnote{In particular, we do not assume a particular spatial
	  dimension.}.

A certain ambiguity is unavoidable at this point,
until a classical description of GR is not chosen. 

Notice that, in the current
 formulations of GR, it is not possible to promote a background with defined metric and
  extrinsic curvature to be a QG state, since they are conjugate variables
   \cite{wald} and so they do not commute.

Recently \cite{botta}, a Yang-Mills-type formulation of GR has been proposed
 for which this is different; the canonical degrees of freedom are $SO(5)$-connections which
  contain information about the metric (vierbein) and its derivatives. The set of possible spatial backgrounds,
characterized by the configurations set of these variables,
 can be promoted to be a "basis of the state space of QG"\cite{wheeler}.

Remark: The form of the most general CSCO in classical physics is $(B,\phi(x))$
 ($x\ep\Sigma_B$). 

The set of CSCO; is the {\it space of degrees of freedom}, referred to as SDF.

When a field theory is quantized (QFT), the background structure is assumed fixed, and the
 theory is determined by a functional $\psi[\phi]$ -the wave function-.
 The set of fields $\phi(x)$, the SDF, constitutes a {\it basis} for the Hilbert space.

 We shall follow the same path to quantize spacetime.
Our main statement is that this space of backgrounds or {\it generalized geometries}
 constitutes a basis for the state space of QG.

Let $\beta$ be the set of generalized backgrounds $B$ \footnote{In the sense described in the
 introduction.}, the SDF of the spatial geometry; then, we {\it promote} this to be
 a basis for the Hilbert pace of Quantum Gravity ($H_{QG}$).

Definition {\bf 2.1}: 
\be
H_{QG}:=\oplus_{B\ep\beta} H_B \label{H}
\ee

Let us motivate this assumption from another point of view, following the same strategy to
 quantize QFT from QM. The structure underneath our construction shall become clear.
\\
\\
\\
{\bf Formal derivation:}

Let us consider a set of $N$ (spatial) points, $B_N$; $B_N \times \R$ is thought of as the basis
 manifold and the field $\phi: B_N \to \R^m$, describes the degrees of freedom at each point in $B_N$.

This system has $N.m$ degrees of freedom; the quantization rules
yield the following Hilbert space:
\be
H_{B_N}=\otimes_{p \ep B_N} H_p ,\label{hil}
\ee
where
\be
H_p=L^2[\R^m].
\ee
Now, consider another set, $B'_{N'}$. We shall have {\it other} Hilbert space with the same
 structure (\ref{hil}), 
$H_{B'_{N'}}$. Notice that, if a bijective map is possible between $B_N$ and $B'_{N'}$ (in the
 discrete case, if $N=N'$); then, they are identified and characterized by $N$,
  $H_{B_N}=H_{B'_N}=H_{N}$.
So, we are able to build a total space: 
\be
H=\oplus_N H_{B_N}.\label{hiltot}
\ee
Now, we follow the same procedure as in the QFT formulation of N quantum-mechanical systems; we may
 consider the continuum-limit: $N \to \infty$, $B_N \to B$: this structure is preserved, and
  $H_B$ is the current Hilbert space for the field $\phi$
on a background $B$. Notice that the isomorphism condition above must be replaced
 by the corresponding equivalence relation: for example, if the metric is one of those
  "background-variables", $H_B$ is defined {\it modulo isometries} \footnote{This depends
   on the classical formulation of GR.}. Thus, we recover structure (\ref{H}), $H \to H_{QG}$.

The similarities between this structure and that of $N$-particles is clear. Then, we can define
 an analogous to the Fock-Space of QG,  ${\cal F}_{QG}\subset H_{QG}$. Let $B_{(n)}$ be a M-M, a collection
  of $n$ {\it non-intersecting}
 manifolds $B_i$; then, the structure of the Hilbert space is such that $H_B \sim \otimes_i H_{B_i}$.
  We define this special subset of quantum states as:
\be
{\cal F}_{QG}=\oplus_n \otimes_{i=1}^n H_{B_i},
\ee
where $n=0$ is included and describes the {\it vacuum state}, which means a no-background
 state \footnote{Another illuminating definition will be discussed in the next section: the "Membrane"
  Hilbert space.}.

 Finally, let us remark that two inequivalent background states are said to be orthogonal, and 
 then the scalar product in $H_{QG}$ naturally is defined from the above considerations.

Finally, we take useful working assumptions in order to have well-defined states and operators in QFT. 
Let us denote by $\phi$, the CSCO of a FT. Then:
\\
\\

{\bf Assumption I:} For every {\it local}\footnote{A pointwise function.} operator $A$ of the theory (FT), we can
 write:
\be
A=\oplus_{B\ep\beta} A_B \left|B\right>\left<B\right|;\label{A}
\ee
where, 
\be
A\left|B\right>=\left|B\right> A_B .\label{AB}
\ee
\\
\\

{\bf Assumption II:} The QFT-wave function is
\be
\left|\psi(\phi,t)\right>=\Sigma_{B\ep\beta} \left|B\right> \left|\psi_B (\phi,t)\right>, \label{AB} \label{psiB}
\ee
where the subscript $B$ denotes, in both cases, the object corresponding to the FT
 {\it found on} the fixed background $B$.
\\
\\

Up to now, we have established an important starting point for the fundamental
 kinematical structure of QG. Remarkably enough, this is {\it background independent}.


There are some questions in QG related to the particular dynamics chosen for GR. But their
 answers are absolutely independent of the ideas exposed above.
  
They are mainly:

1-What are the B-variables of the theory?, 

2-They can be a full (spatial) background geometry (topological structure, metric,
 connection/covariant derivative), a state of QG?.
 
  Finally,

3-What are the quantum equations for GR?

Remarkably enough, notice that: "requeriments for these answers may be deduced in order
to have a  well-defined QFT for $\phi$". 

Some quantities related to the extrinsic geometry of $set(B)$, {\it extrinsic
 variables}\footnote{On the manifold $set(B)\times\R$.}, -but valued on this surface-
could also be required to have a well-defined QFT.

\section{p-brane states.}
An essential property of the membrane fashion is the "ambient space" or target. From our
 point of view (background independent), we do not start off with this: the fundamental objects
 are the "generalized backgrounds". The main question of this section is: How can the notion
 of ambient space arise in this framework?

Now, we discuss more deeply the structure of $H_{B}$
in order to show that this language can serve to describe membranes and string states. Let us
 denote by $a/b$ the set of functions from the set $b$ in the set $a$, and redefine the
  standard notation as follows: $L^2[C/K]:=L^2[K]$\footnote{It denotes formally the set
   of square-integrable complex functions on $K$, suposing that this exists.}.
If the degrees of freedom of a {\it full} QFT (including gravity) are summarized by $(B,\phi)$,
 then we can write, 
\be
H_{B}=\{\left|B\right>\}\otimes H_{\phi},
\ee
 where $H_{\phi}$ denotes the usual Hilbert space in QFT for the fields $\phi$. If 
 $\phi:B\to \Phi$\footnote{When there are not fields $\phi$, this is a one-dimensional space, and
 the theory is an enterely geometrical one.}, then its structure is
   $L^2[C/(\Phi/B)]$\footnote{ Recall that we build up the Hilbert space using the canonical
    rule: $H=L^2[C/(SDF)]$, where SDF is {\it the space of degrees of freedom}.}. 

Actually, as we have argued in the first section, if $B$ is a one-component manifold,
 \footnote{It is not a multi-manifold.} once the atlas is given
  (i.e the set of the points of $B$ is specified ), we shall denote this by $set(B)$;
   then, all the local $B$-variables together with  the rest of the fields; these are fields on
    $set(B)$, valued in some manifold ${\cal F}$ (the fiber).  Thus, the full structure of the
	 Hilbert space for a background can be written,
\be
H_{B}=\{\left|set(B)\right>\} \otimes L^2 [C/({\cal F} / set(B))].\label{intHB}
\ee


Then, we define a {\it sub-background} or p-brane state (p is the spatial dimension) $\left|B\right>$, if:

 There exists a decomposition  ${\cal F}= {\cal A} \times \Phi$;  ${\cal A}$ 
   is called {\it ambient} space or target-, such that :

{\bf I-} ${\cal A}\ep \beta$\footnote{Then, ${\cal A}$ has geometry-variables too. Besides that, there would
 be fields defined on ${\cal F}$ , further to the gemetrical ones.}.

{\bf II-} The full {\it local} structure of $B$ ($loc(B)$) can be induced from those of
 ${\cal A}$, via an element $x \ep {\cal A} / set(B)$. It can be written
\be
\left|loc(B)\right>=\left|{\cal A},x\right>.
\ee
In general, we can decompose the B-variables as
\be
\left|B\right>=\left|set(B)\right>\left|loc(B)\right>
\ee
 thus,
\be
\left|B\right>=\left|set(B)\right>\left|{\cal A},x\right>.\label{SUB}
\ee

{\bf III-} For the structure (\ref{intHB}), $x$ is a physical degree of freedom -it
 is a dynamical
 field-\footnote{In particular, this must be quantized.}.

Let us observe that non-scalar embedding fields $x$, the target ${\cal A}$ is
 a so-called non-commutative geometry.

An important remark: notice that the embedding in a "major" manifold is generically possible;
 but the main point which characterizes a "sub-background physical state" is that the embedding
  field $x$ is a {\it physical degree of freedom}, and thus in particular it must be
   quantized.

This definition can be heuristically derived in a similar form to those of the first
 section, supposing subsets $B_N$ in a "major" continuum background $M$.

Notice that this definition contains in some sense the intuitive idea of these objects as "distributional" ones.
Since ${\cal A}\ep \beta$, then in principle, we could write {\it formally} an alternative to
 the expression (\ref{intHB}):
\be
H_{B}=\left|{\cal A}\right> \otimes "L^2 [C/({\cal F} /set({\cal A}))]"\label{dist}
\ee
The $""$ expresses that here $({\cal F} / {\cal A})$ must be substituted by a more general
 set; the set of "generalized" functions on ${\cal A}$ -distributions-, and then $L^2$ should
  be replaced by some corresponding Sobolev space. Then, it is clear that the "sub-branes"
   can be described distributionally as  intuitively expected; however, it seems more
    convenient to adopt structure (\ref{intHB}), that is unique.
\\
\\

{\bf Intersecting sub-backgrounds:}
\\
\\

Consider $B=B_1 U B_2$, such that
\bea
\left|B_1\right>=\left|set(B_1)\right>\left|{\cal A},x_1\right>,\\
\left|B_2\right>=\left|set(B_2)\right>\left|{\cal A},x_2\right>
\eea

Now, we say that $B_1 , B_2$, are intersecting if and only if the functions $x_1 , x_2$ coincide
 at some point. Then, let a configuration given by
 $\left|{\cal A},x_1,x_2 : x_1(p_1)=x_2(p_2) \right>$ -where $p_{1/2}\ep B_{1/2}$-; 
\be
\left|\psi(x_1,x_2)\right|^2=\left|<{\cal A},x_1,x_2 : x_1(p_1)=x_2(p_2)\left|\psi\right>\right|^2
\ee
gives the probability that $B_1 , B_2$, intersect at one point $p:= p_1 = p_2 $.

 As a precise example of this discussion, we write down the Hilbert space of the states of a string theory:
\be
set(B)=[0,1]
\ee
\be
H_{1-string}=\left|[0,1]\right> \otimes L^2[C/({\cal T}/[0,1])].
\ee
where ${\cal T}$ is a Riemannian 10-dimensional manifold known as target space (we are concerned with the
 bosonic sector). The Hilbert space of the states of string theory is:
\be
H_{strings}=\oplus_n [\otimes^n H_{1-string}].
\ee

 Notice that this has the same structure of a Fock-type structure (${\cal F}_{QG}$)
  defined in the first section.

 Closed string-space is expanded by the basis elements:
\be
\left|{\cal S}\right>:=\left|[0,1]\right> \otimes \left|{\cal T}\right> \otimes \left|x(s):x(0)=x(1)\right>
\ee
 Then,  the wave functions: $\left<\psi|{\cal S}\right> =\psi[x(s)]$.
\\
\\

{\bf Loops:}
\\

Another non-perturbative approach to quantum gravity recently developed \cite{loop} is based
 on geometrical one-dimensional structures embedded in a three-dimensional one,
namely: loops. Also, they can be described in this language:

The ambient space is a three-dimensional (compact) manifold, $\Sigma$, and the "backgrounds"
 correspond to
$S^1$, or a collection of circles:
\be
set(B)=S^1 .
\ee
Finally, the B-variables are induced by those of $\Sigma$ via the embedding
 $\gamma^{\mu}:S^1\to\Sigma$;
these variables, first introduced by Ashtekar \cite{ash}, are the 3-dimensional space vierbeins $E$ or
 their canonical conjugate:
the $SU(2)$-connection $A$. 

Then, a loop state is $\left|\lambda\right>=\left|S^1\right>(\left|A ; \gamma^{\mu}\right>)$, which resembles
 (\ref{SUB}).

According to our construction, these are the elements which we need to express the
 Hilbert space.
The loop-QG Hilbert Space is
\be
H_{loop}=\oplus_{n=0}[\otimes^n [\left|(S^1)\right>\otimes L^2(C/(\Sigma/S^1))]].
\ee
 Then, a state in the $A$-basis takes over the form $\psi_{\gamma^{\mu}}[A]$. 

\section{Concluding remarks.}

Our claim is that the present work is the most general fashion  for the space (or space-time)
 at
 {\it observable level} -direct or indirectly-. That is to say: if there are more fundamental
  structures, they must have {\it contact} with observable ones; which we believe to be the
   ones, proposed in this work.
In a future work, following up this one, we exploit the dynamics due to new
 formulation of GR \cite{botta}
and construct a particular, interesting full-model for QG, which complements the main ideas
 of this work.
 
 The possibility to describe, as it has been shown in the final examples,
 the kinematical objects of the recent more promising approaches to a fundamental theory in 
 terms of the concepts formulated in this work, such as p-branes, strings, loops, and others,
 allows us to argue that they are simply "subspaces of our $H_{QG}$".

 Finally, we hope that a diagrammatics for evolving $B$-geometries will arise when
 intersections/interactions are considered. In a "Feynman-rule"
picture, where the time development of $B$ would agree with a Feynman-diagram \cite{pst}.
 The recent Spin Foam Models seems to be examples of this. See for instance \cite{rr}.

\section{Acknowledgments.}

The author expresses his gratitude to Prof. Carlo Rovelli for helpful and
 very pertinent comments.
Thanks are due also to J. A. Helayel-Neto for encouragement and a critical reading
 of the manuscript in its earlier form. CNPq is also
acknowledged for the invaluable financial help.

\end{document}